# Symmetry-Gated Sliding: Decoupling Interlayer Translation from Polarization Reversal in Boron Nitride Ferroelectrics

*Yinan Wang*[1,2], *Teruyasu Mizoguchi*[2*]

[1]Department of Materials Engineering, the University of Tokyo, Tokyo, Japan

[2]Institute of Industrial Science, the University of Tokyo, Tokyo, Japan

**ABSTRACT:** Electric-field-driven polarization switching in sliding ferroelectrics depends on how local interlayer motion is accommodated by the surrounding structure. Yet the relationship between sliding, domain-wall (DW) propagation, and collective domain evolution under changing vector fields remains unclear. Here, we show that local symmetry governs the coupling between field-induced forces and interlayer sliding, while DW connectivity determines how local motion contributes to net domain conversion. These effects are examined across three structural scales, from monodomains and isolated DWs to reconstructed moiré networks. We investigate bilayer hexagonal boron nitride (h-BN) using molecular dynamics with a machine-learning interatomic potential that incorporates field-dependent electric enthalpy. An initially monodomain bilayer switches through directional wrinkling and interlayer sliding, while the barrier along a constrained flat path remains finite. Pre-existing DWs propagate with velocities governed by their shear and tensile character, and an additional in-plane field tunes wall velocity and width according to its orientation. In reconstructed triangular moiré networks, shared junctions and boundary curvature mediate domain evolution and recovery after field removal or reversal. Most strikingly, a rotating in-plane field translates the layers and displaces the moiré nodes by approximately 1 Å yet leaves the AB/BA domain-area ratio without detectable change, because threefold connectivity cancels the contributions of individual walls. We term this response polarization-silent sliding. These results distinguish local sliding, DW translation, and net domain conversion as different responses to electric fields and establish how local structural symmetry and DW connectivity shape their selective control.

## INTRODUCTION

Sliding ferroelectricity provides a distinct mechanism for controlling electrical polarization in van der Waals materials through relative translation between adjacent atomic layers.[1–7] In parallel-stacked bilayer hexagonal boron nitride (h-BN), the AB and BA configurations possess opposite out-of-plane polarizations arising from interfacial charge redistribution. Their interconversion through lateral sliding directly couples polarization reversal to structural motion.[1,5,8] Such stacking-dependent polarization switching has been experimentally established across several van der Waals bilayers, including transition metal dichalcogenides, and has also been resolved optically in h-BN-based heterostructures.[9–12] This coupling has enabled ultrafast and fatigue-resistant ferroelectric devices and established interlayer sliding as a promising route toward electrically controlled functionalities in atomically thin materials.[2,8] In realistic bilayers, however, polarization reversal involves spatially nonuniform stacking configurations and the motion of boundaries separating oppositely polarized domains.[13,14] Understanding how these local structural motions produce collective polarization switching is therefore central to the dynamics of sliding ferroelectrics.

The microscopic mechanism of field-driven sliding is closely connected to the symmetry of the local stacking environment. In an ideal AB or BA monodomain, threefold rotational symmetry prevents an out-of-plane electric field from selecting a unique in-plane sliding direction.[6] Breaking this symmetry enables off-diagonal components of the Born effective charge (BEC) tensor to generate the lateral forces required for interlayer motion.[13,15] Recent calculations have demonstrated that symmetry breaking through uniaxial strain can enable directional sliding even in monodomain structures.[16] In multidomain systems, domain walls (DWs) naturally provide lower-symmetry environments where local stacking rearrangements propagate polarization reversal.[17–19] Recent experimental measurements have further revealed in-plane polarization localized along the DWs of twisted h-BN, highlighting their distinct local polar structures.[20,21] Other experimental imaging has directly resolved polarization switching through DW motion in rhombohedral $MoS_2$, while local electric-field manipulation of twisted $WS_2$ has demonstrated reversible deformation and reconfiguration of DW networks.[22,23] Atomistic simulations have revealed coherent, wavelike DW motion with exceptionally high propagation speeds and switching fields substantially below those associated with homogeneous reversal.[13,24,25] Furthermore, DWs with different armchair and zigzag geometries exhibit distinct structural profiles and switching behavior.[13,24,26] These findings establish local symmetry and DW structure as essential ingredients of electric-field-driven polarization conversion.

The situation becomes more complex when DWs form connected networks. Small-angle twisting produces reconstructed moiré superlattices containing alternating polar domains separated by extended DWs and interconnected junctions.[5,14,26–28] Experimental studies of twisted h-BN have revealed complex polarization textures within these networks, including in-plane polarization along domain boundaries and additional out-of-plane polarization associated with local reconstruction and piezoelectric effects.[20,21] Unlike an isolated wall, the displacement of one boundary in such a network must be accommodated by neighboring segments and their shared nodes. Operando electron microscopy of twisted $WSe_2$ has demonstrated that DW network topology can prevent the conversion of alternating polar domains into a uniformly polarized state, even when individual boundaries respond to an applied field.[29] Theoretical studies have likewise shown that moiré geometry influences domain evolution, polarization retention, and recovery after field removal.[13,26] These observations highlight the importance of collective structural constraints in determining the macroscopic electric response. However, these studies have focused on out-of-plane fields, and it remains unknown whether network constraints merely impede domain conversion or can decouple structural motion from polarization altogether.

Although the role of DWs in polarization reversal is well established, the relationship between local sliding dynamics and collective domain conversion remains less clear. Electric fields can drive multiple coupled structural responses, including interlayer translation, changes in wall position and width, and deformation of connected junctions. These motions need not produce equivalent changes in polar-domain occupation. Moreover, the response to an applied field depends on both its direction relative to the local sliding geometry and the structural constraints imposed by neighboring domains.[30] A central question is therefore how structural symmetry, DW geometry, and network connectivity determine whether, and to what extent, electric-field-driven interlayer sliding develops into collective ferroelectric switching. Answering this question would turn the distinction between sliding and switching from an empirical observation into a design principle. It would identify which structural features must be broken to lower the switching field, and which must be preserved to move the lattice without disturbing the stored polarization. A single electric field could then serve two distinct functions: writing polarization or actuating the structure while leaving the stored state intact. Here, we investigate this relationship in bilayer h-BN using molecular dynamics with a machine-learning interatomic potential that explicitly incorporates the coupling between electric fields and atomic configurations through the electric enthalpy.[31] We analyze the response in terms of three distinct coordinates: the relative interlayer translation, the DW position, and the domain-area ratio $A_{AB}/A_{BA}$, which measures net polarization reversal. An applied field does work on these coordinates only through its projection on the corresponding structural modes, and symmetry constrains this projection. We examine initially monodomain structures, four distinct DW geometries, and reconstructed triangular moiré networks under out-of-plane and inclined electric fields. In monodomains, where $C_{3v}$ symmetry forbids a net sliding force at the exact AB registry, polarization reversal accompanies directional wrinkling that breaks the initial symmetry and enables sliding while the barrier along a constrained flat pathway remains finite. In isolated DWs, the reduced $C_s$ symmetry permits field-sliding coupling, and the local shear and tensile character governs propagation speed, while an additional in-plane field produces structure-dependent angular responses in wall motion and deformation. Connecting these walls into a moiré network introduces collective geometric constraints that govern domain evolution and recovery following changes in the applied field. Notably, in-plane fields can induce interlayer sliding and geometrically amplified node displacements without appreciably changing the relative areas of oppositely polarized domains, because the contributions of individual walls to the total area cancel to first order in a threefold network. We refer to this response as polarization-silent sliding. It provides a direct example of a field that actuates the structure while leaving the stored polarization intact, which extends previous observations of domain evolution under out-of-plane fields by revealing an additional structural response to an in-plane field.[3,13,19,26,29] Together, these results show that electric-field control of sliding ferroelectrics is gated twice. Local symmetry determines whether the field can drive sliding, and domain connectivity determines whether sliding becomes switching.

## RESULTS AND DISCUSSION

### Field Coupling and Structural Degrees of Freedom in Monodomain Switching

We first examine the field-driven evolution of an initially AB-stacked monodomain bilayer. In the absence of pre-existing DWs, this system allows us to follow how structural deformation and interlayer sliding develop into AB-to-BA switching. We simulate field-driven dynamics using a machine-learning interatomic potential. Figure 1A illustrates the two contributions to out-of-plane polarization. The AB and BA registries exhibit opposite spontaneous ferroelectric polarizations arising from interfacial charge redistribution, which can be reversed through interlayer sliding. An applied out-of-plane field also induces opposite vertical displacements of the B and N sublattices within each layer, generating an additional ionic polarization. This contribution develops without changing the interlayer registry and therefore affects the polar signal within each stacking branch.

The monodomain field sweeps in Figure 1B reveal AB-to-BA switching under a strong negative out-of-plane field and the reverse transition on the opposite sweep branch. The background-corrected polarization signals show switching near $|\varepsilon_z| = 0.45$ V Å$^{-1}$, with similar switching intervals across the sampled temperatures. Switching is also observed across different initial-velocity seeds. These switching fields characterize the present finite-time and finite-cell simulations.

To understand the observed switching behavior, we consider the symmetry and field-dependent energetics of the initial AB structure. According to Neumann's principle, the response of an ideal $C_{3v}$-symmetric bilayer must preserve its threefold rotational symmetry.[13,32] A net in-plane force along a specific sliding direction would break the equivalence of the three symmetry-related directions. Consequently, an out-of-plane field cannot produce a net driving force along the collective sliding mode at the exact AB registry. However, coupling between the field and stacking-dependent polarization reshapes the electric enthalpy landscape, changing the relative stability of AB and BA and reducing the barrier for switching. We first examine this energetic effect along a constrained flat sliding path and compare it with the switching behavior observed in MD.

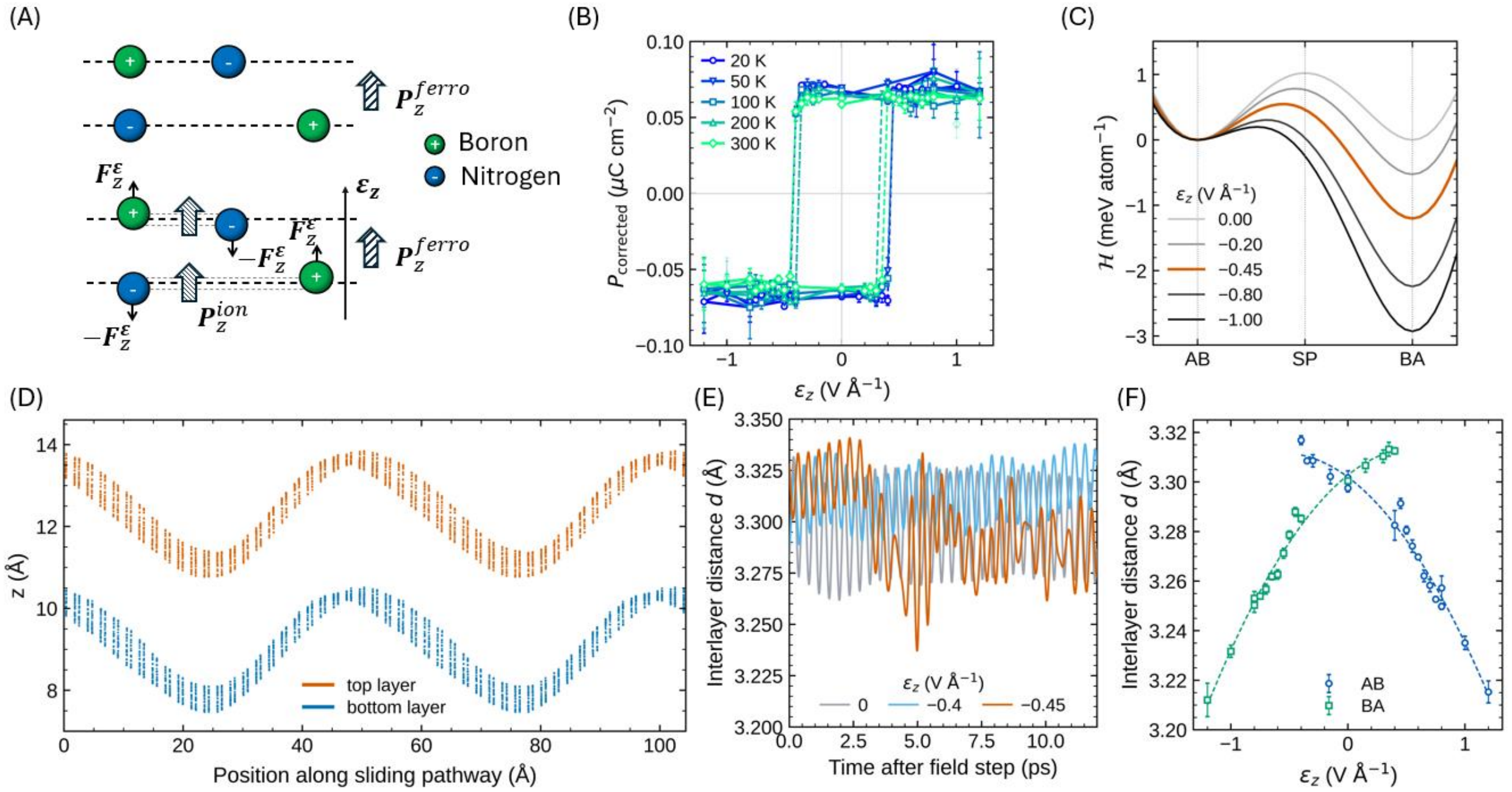

Figure 1. Field coupling and structural evolution during monodomain switching. (A) Schematic of ferroelectric polarization and field-induced ionic polarization arising from vertical B–N sublattice displacements. (B) Background-corrected polarization signals during electric-field sweeps at 20–300 K. Details of the polarization processing for the hysteresis loops are provided

in Section S3 of the SI. (C) Field-dependent energy profiles along the flat AB-to-BA sliding path. (D) Directional wrinkling along the sliding direction immediately before switching at −0.45 V $Å^{-1}$. Orange and blue represent the upper and lower layers. (E) Evolution of the interlayer distance following field steps to 0, −0.40, and −0.45 V $Å^{-1}$. (F) Field-dependent interlayer distance for AB and BA stacking, showing opposite responses under the applied electric field.

Figure 1C shows the electric enthalpy profiles along the constrained flat sliding path under different out-of-plane fields. Increasing the negative field stabilizes BA relative to AB and lowers the intervening energy barrier. The saddle-point (SP) barrier vanishes near εz = −0.82 V $Å^{-1}$. In contrast, switching occurs in MD at εz = −0.45 V $Å^{-1}$, where the barrier along the flat path remains finite. The MD trajectory also exhibits a directional wrinkle before switching, introducing a structural deformation absent from the constrained flat path (Figure 1D).

The pre-switching structure identifies a directional wrinkle as this additional degree of freedom. At $\varepsilon_z$ = −0.45 V $Å^{-1}$, 0.02 ps before the identified SP crossing, both layers exhibit a common height modulation with a wavelength of 52.2 Å and an approximate peak-to-trough excursion of 2.5 Å. Their correlated profiles show that the bilayer bends coherently while retaining a much smaller variation in the local layer separation. The oriented height modulation lowers the instantaneous structural symmetry below $C_{3v}$ and selects an in-plane sliding direction.

In the distorted structure, symmetry permits off-diagonal BEC components to couple the out-of-plane field to in-plane atomic motion. The corresponding field-induced forces can acquire a nonzero projection along the selected interlayer sliding mode. Within MACE-Field, this response is included through the coordinate dependence of the field-dependent electric enthalpy. The projected force histories in Figure S2 are consistent with this mechanism. The in-plane driving force increases before switching, while the restoring force becomes insufficient to maintain the initial registry, eventually leading to AB-to-BA conversion. Furthermore, the approximately quadratic increase in ripple intensity with field magnitude indicates a field-even deformation response. Together, the trajectory and force histories support a wrinkling-assisted route that becomes accessible before the barrier on the constrained flat path disappears.

Interlayer breathing accompanies this reconstruction. Figure 1E compares zero field, a non-switching segment at −0.40 V $Å^{-1}$, and a switching segment at −0.45 V $Å^{-1}$. The distance traces occupy approximately 3.24–3.34 Å and exhibit oscillations around field- and registry-dependent mean values. The switching segment evolves toward a smaller mean distance, with a changed oscillation envelope as the bilayer enters the opposite registry. Across the full sweep (Figure 1F), the AB and BA distance branches approximately map onto each other under simultaneous registry interchange and field reversal. This branch relation allows a linear response within a fixed polar branch and an approximate symmetry between the paired branches.

Monodomain switching under strong out-of-plane fields involves both a reduced electric enthalpy barrier and directional wrinkling that breaks C3v symmetry, enabling interlayer sliding while the constrained flat-path barrier remains finite. In contrast, pre-existing AB-BA DWs provide lower-symmetry environments for field-driven switching at substantially weaker fields. Within the framework introduced above, the monodomain represents the symmetry-forbidden limit. The field cannot act on the relative layer translation until the structure itself breaks C3v, which explains why switching requires a field an order of magnitude larger than those needed for wall motion.

## Structure-Dependent Propagation of Four Ferroelectric DWs

Pre-existing DWs provide a local sliding pathway for polarization reversal at fields well below those required for the initially monodomain bilayer. We examine four DW configurations, including two armchair (AC-I and AC-II) types and two zigzag (ZZ-I and ZZ-II) types in periodic rectangular cells containing a central AB domain between two BA domains (Figure 2A).

After 10 ps of relaxation at zero field, applying $\varepsilon_z = -0.03$ V Å$^{-1}$ causes the AB domain to contract in all four cells (Figure 2D). The propagation velocity follows AC-I ≫ ZZ-I > AC-II > ZZ-II (Figure 2B). AC-I reaches approximately 13 Å ps$^{-1}$ and consumes the AB domain after about 20.6 ps of field exposure, while the other walls move below 1 Å ps$^{-1}$. The different propagation rates can be understood from the local sliding geometries. AC-I and ZZ-II have pure shear and tensile character, respectively, while AC-II and ZZ-I contain different mixtures of these components. The speed ordering follows the progression from shear to tensile sliding. The slower walls also broaden more strongly, whereas AC-I retains a comparatively narrow profile during propagation.

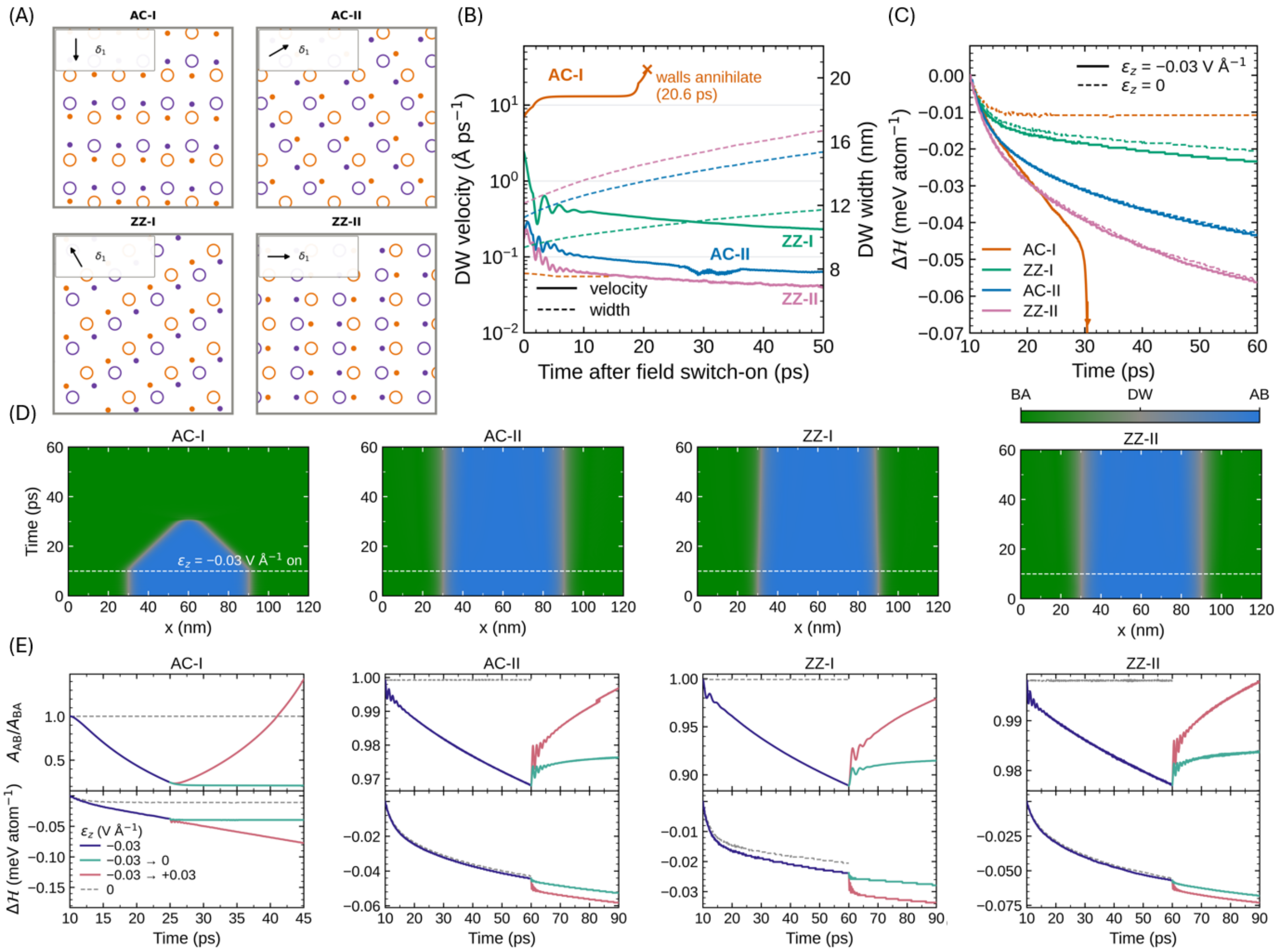

Figure 2. Structure-dependent propagation of four ferroelectric DWs and their response to electric field removal and reversal in bilayer h-BN. (A) Atomic configurations of AC-I, AC-II, ZZ-I, and ZZ-II walls. Arrows indicate the relative interlayer displacement vectors associated with registry conversion. (B) Wall velocity and width versus time after application of $\varepsilon_z = -0.03$ V Å$^{-1}$. Solid curves show velocities on the logarithmic left axis, and dashed curves show widths on the right axis. (C) Electric enthalpy changes per atom under $\varepsilon_z = -0.03$ V Å$^{-1}$ and at zero field, shown by solid and dashed curves, respectively. (D) Registry maps as functions of position and time. Blue and green areas indicate AB and BA domains, respectively. White dashed lines mark field application at t = 10 ps. (E) AB/BA area ratio and electric enthalpy changes per atom during field removal and reversal. The initial negative field is changed to zero or +0.03 V Å$^{-1}$ at t = 25 ps for AC-I and t = 60 ps for the other structures. Blue curves show the initial negative field interval. Turquoise and pink curves show subsequent evolution after field removal and reversal, respectively. Gray dashed curves denote controls maintained at zero field throughout.

The electric enthalpy histories reveal how strongly the applied field contributes to the observed motion (Figure 2C). AB and BA have equivalent zero-field energies. In MACE-Field, electric enthalpy $\mathcal{H}$ is expressed as

$$\mathcal{H} = E_0 - \Omega \boldsymbol{\varepsilon}_z \boldsymbol{P} \tag{1}$$

Here, $E_0$ denotes the zero-field total energy evaluated at the instantaneous atomic configuration. During wall propagation, the changing domain fractions alter the net polarization and its coupling to the applied field. For AC-I, the zero-field structural energy changes comparatively little before annihilation, while the coupling between polarization and field lowers the electric enthalpy as the walls move. This accounts for the substantial separation from its zero-field control. When the two walls meet and annihilate, an additional sharp decrease accompanies the loss of the interfaces. ZZ-I also separates visibly from its zero-field reference. For AC-II and ZZ-II, the curves under the applied field remain close to their zero-field controls despite a continued decrease in energy. Structural relaxation therefore accounts for much of the energy evolution in these slower cells. The comparison explains why similar downward energy trends can accompany very different amounts of domain conversion.

We next change the field after the walls have propagated under $\varepsilon_z = -0.03$ V Å$^{-1}$ (Figure 2E). The field is changed at t = 25 ps for AC-I, before the walls annihilate, and at t = 60 ps for the other structures. From the same configuration at this time, one trajectory continues at zero field and another receives $\varepsilon_z = +0.03$ V Å$^{-1}$. These two branches show whether domain contraction persists without the original driving field and whether reversing the field can reverse the wall motion.

Removing the field nearly arrests the contraction of the AC-I domain. After a short adjustment, its AB area fraction approaches a plateau, and its electric enthalpy also becomes nearly constant.

The slower structures show a small recovery of AB area after removal, while their energies continue to decrease as the cells relax. Their zero-field references in Figure 2C likewise continue to lose energy over this interval. The sustained energy decrease after removal is thus consistent with ongoing relaxation, while the domain area records the much smaller change in wall position.

Reversing the field produces a qualitatively different response. The AB area increases in all four cells, showing that the walls reverse their direction of propagation. The effect is most pronounced for AC-I, whose AB fraction rises from approximately 19% at the field change to about 59% at t = 45 ps, exceeding its initial value of 50%. The slower walls also reverse their area trends, with a larger recovery under the positive field than after field removal. The rapid adjustment of the AB area fraction at the field change in AC-II, ZZ-I, and ZZ-II is absent for the purely shear AC-I wall. This contrast may reflect rapid strain relaxation associated with the tensile components of these walls before sustained reverse propagation develops. Thus, changing the sign of the out-of-plane field changes which polar domain expands, while the wall structure determines how rapidly the conversion proceeds.

Figure 2E explains how motion in either direction can be favorable. Under the initial negative field, the electric enthalpy decreases as BA expands. After reversal, the energy continues to decrease while AB expands and the walls move back toward their original positions. The field changes the energetic preference between the two registries, allowing the direction of domain conversion to reverse while energy is released along the new trajectory. The abrupt energy change at reversal is followed by a further decrease during wall motion. At a given domain area, the energy depends on both the applied field and the relaxed atomic structure. AC-I consequently passes through its initial area fraction at a lower electric enthalpy during the reversed trajectory.

This response demonstrates the importance of describing the coupling between the electric field and the evolving structure consistently in both energies and forces. In MACE-Field, the electric enthalpy depends explicitly on the applied field, so reversing the field changes the energetic driving force for domain conversion. The combined area and energy histories establish control of the propagation direction through the sign of the out-of-plane field. These walls therefore represent the symmetry-permitted limit. Because the reduced symmetry opens a finite field-sliding projection, the wall geometry, through its shear or tensile character, sets how efficiently that projection is converted into wall translation. We next examine how an additional in-plane field changes the propagation speed for a fixed out-of-plane bias.

**Angular Control of DW Translation and Deformation**

The angular response of a ferroelectric DW connects its propagation to the deformation of the surrounding lattice. With the out-of-plane bias fixed, we rotate a weaker in-plane field relative to each prepared wall and compare the resulting velocity, registry transition width, and B/N relative displacement. This common field protocol reveals which aspects of the response depend on wall structure and whether the field direction that accelerates translation also produces the largest structural change. The comparison extends the control of propagation direction established in Figure 2 to control of the rate and geometry of the response.

The angular series uses $\varepsilon_z = -0.03$ V Å$^{-1}$ and an in-plane amplitude $\varepsilon_{\parallel} = 0.01$ V Å$^{-1}$, with $\theta$ sampled every 30°. In each cell, $\theta = 0°$ points along the wall normal +x, and $\theta = 90°$ lies along the wall tangent +y. The in-plane field components are therefore $\varepsilon_{\parallel} \cos \theta$ and $\varepsilon_{\parallel} \sin \theta$. The prepared wall and crystal orientation remain fixed during the angle scan, so rotating the field directly probes the directional response of that interface.

The AC-I field-amplitude series shows that the in-plane field also changes the wall width (Figure 3A). After the initial oscillations, the width approaches a broader profile as positive $\varepsilon_y$ increases. The auxiliary field therefore changes the spatial extent of the registry transition while the out-of-plane field drives its propagation. Furthermore, positive $\varepsilon_y$ advances wall-pair annihilation and negative $\varepsilon_y$ delays it (Figure 3B). AB contraction persists across the displayed amplitude series, including positive $\varepsilon_y$ up to 0.03 V Å$^{-1}$. Within this range, the auxiliary field adjusts the conversion rate while the fixed out-of-plane component maintains the direction of domain growth. The propagation direction is therefore governed primarily by the out-of-plane field, with the in-plane component providing an additional means of control.

Figure 3C compares the relative velocity change, expressed as a percentage, for the four wall types. Filled and open markers denote acceleration and deceleration relative to the baseline without the in-plane field, respectively.

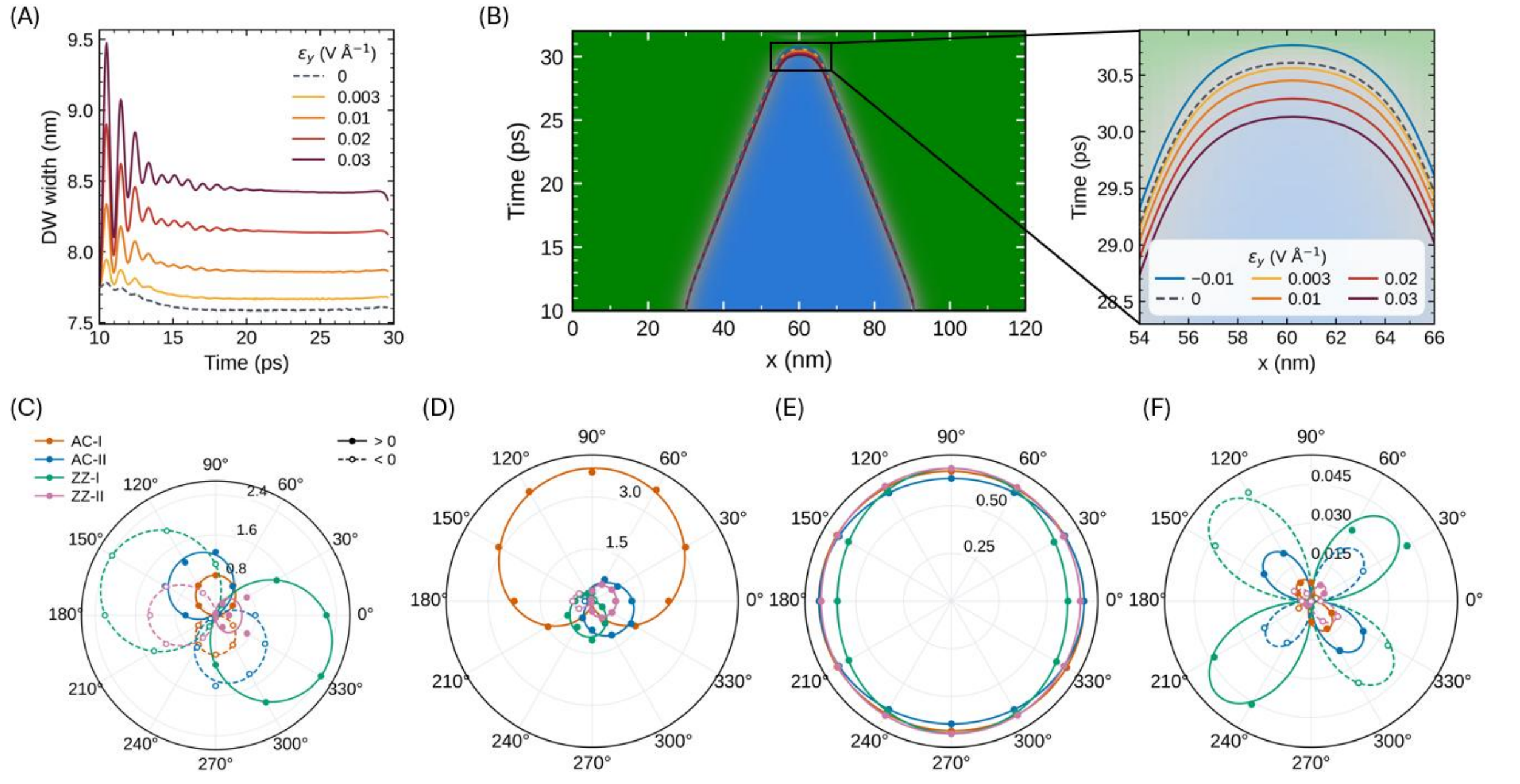

Figure 3. Angular control of DW translation and structural response under an inclined electric field. (A) Time evolution of the AC-I wall width for different positive $\varepsilon_y$ at fixed $\varepsilon_z = -0.03$ V Å$^{-1}$. The dashed curve denotes $\varepsilon_y = 0$. (B) AC-I registry evolution and an enlarged view near wall-pair annihilation for the indicated $\varepsilon_y$ values. (C, D) Relative changes in wall velocity and width, respectively, expressed as percentages of their values without the in-plane field. Velocities are obtained from linear fits to the wall positions and averaged over the two walls. Widths are evaluated at 20 ps using equivalent tanh profiles. (E, F) Changes in the B/N sublattice centroid separation projected parallel and perpendicular to the in-plane field, respectively, averaged over both layers. Panels C to F use $\varepsilon_z = -0.03$ V Å$^{-1}$ and $\varepsilon_\parallel = 0.01$ V Å$^{-1}$, with θ measured in 30° steps. All changes are relative to the corresponding trajectory with $\varepsilon_\parallel = 0$. Filled and open symbols indicate positive and negative changes, with radial distances showing their magnitudes. Orange, blue, green, and magenta identify AC-I, AC-II, ZZ-I, and ZZ-II, respectively.

AC-I and ZZ-I exhibit a clear difference in their preferred field directions. AC-I reaches its largest positive velocity increment near θ = 90°, where the in-plane field is parallel to the wall tangent and the principal local sliding direction. By contrast, ZZ-I exhibits a larger fractional modulation, with its maximum among the sampled angles at θ = 330°. This direction combines a positive normal component with a negative tangential component in the local wall frame.

Nevertheless, AC-I propagates much faster than ZZ-I under the out-of-plane field, so a smaller percentage change can still produce a larger absolute velocity increment. The change in angular maximum shows that the effective direction for controlling propagation depends on the wall structure. Meanwhile, AC-II exhibits a small absolute modulation because of its much slower baseline propagation. The ZZ-II velocity change is comparable to the reported numerical noise, and its apparent polar lobes provide little evidence for a preferred propagation direction.

Taken together, these observations indicate that the in-plane field provides a direction-dependent modulation of DW propagation under an out-of-plane driving field. Its effect therefore depends on the field direction relative to the wall and its local sliding path. More specifically, the nearly tangential maximum of AC-I is consistent with its shear geometry, while the oblique maximum of ZZ-I reflects a different collective response. The $C_s$ symmetry of the reference wall region is consistent with different responses along the wall normal and tangent.

The angular width response is strongest for AC-I, which exhibits its largest broadening when the in-plane field aligns with the local atomic sliding direction at $\theta = 90°$ (Figure 3D). Notably, fields along the wall normal at $\theta = 0°$ and $180°$ produce much smaller changes. This angular dependence indicates that wall broadening is primarily driven by field-assisted atomic sliding. As atoms within the wall slide, neighboring atoms on both sides are drawn into the registry rearrangement, extending the transition region and increasing the wall width. In comparison, direct stretching or compression of B–N bonds contributes less to the width change. The response to field reversal further distinguishes AC-I from the other walls. A field opposite to the sliding direction leaves the AC-I width nearly unchanged, whereas the other wall types exhibit narrowing under reversed fields despite their similar broadening tendencies under forward fields.

The relative displacement between the B and N sublattices is dominated by the component parallel to the applied in-plane electric field (Figure 3E). This longitudinal response is nearly independent of the field direction and remains similar across the four wall types, consistent with the opposite BECs of B and N. ZZ-I exhibits a slightly weaker response, particularly for fields near $\theta = 0°$ and 180°. In contrast, the much smaller transverse displacement reveals a clear angular dependence (Figure 3F). The transverse component changes sign with field direction, producing a four-lobed pattern. ZZ-I exhibits the strongest transverse response, while AC-II shows a smaller response with the opposite sign at corresponding field angles. AC-I and ZZ-II exhibit only weak transverse displacements. These results indicate that the ionic response is predominantly aligned with the applied field, while the local DW structure introduces a smaller anisotropic component.

An in-plane electric field thus provides an auxiliary means of tuning the velocity and width of a ferroelectric DW through its magnitude and direction. Within the conditions studied, the out-of-plane field remains the primary driver that determines which polar domain expands. The longitudinal ionic response is largely shared by the h-BN layers, while the smaller transverse response resolves the anisotropy of the evolving structures. The angular maps show that, once symmetry permits coupling, the direction of the field relative to the local sliding vector controls its efficiency. Translation, broadening, and ionic displacement each follow different projections. We next examine how these local responses combine in triangular moiré domains, where differently oriented walls are connected through shared nodes.

**Collective Field Response of Triangular Moiré Domain Networks**

Rotating one h-BN layer by 0.336° initially produces a moiré pattern in which AB, BA, and AA registries occupy hexagonal regions of approximately equal area (Figure 4A). During 10 ps of zero-field relaxation, the energetically unfavorable AA regions contract as AB and BA expand, producing the triangular domain network in Figure 4B. Small AA regions remain at the junctions, with approximately 1000 atoms associated with each node in the simulated structure.

The triangular AB and BA domains are bounded predominantly by shear-like walls that meet at AA nodes. These segments share the favorable local sliding geometry of shear walls (AC-I), while their connectivity requires adjacent boundaries and junctions to evolve together. The resulting domain conversion depends on how local wall motion changes the shape and area of the entire triangle.

Starting from the relaxed configuration, separate trajectories receive $\varepsilon_z = -0.03$ or $+0.03$ V Å$^{-1}$ at t = 10 ps. We track the AB/BA area ratio, $R = A_{AB}/A_{BA}$, which initially equals unity. After 60 ps of field exposure, R decreases to approximately 0.70 under the negative field and increases to approximately 1.40 under the positive field (Figure 4C,D,G). The two responses are approximately related by $R(-\varepsilon_z) = 1/R(+\varepsilon_z)$, showing that field reversal exchanges the growth and contraction of AB and BA. This reciprocal relation expresses the symmetry of the domain response. Both branches respond rapidly after field application and then evolve more slowly (Figure 4G).

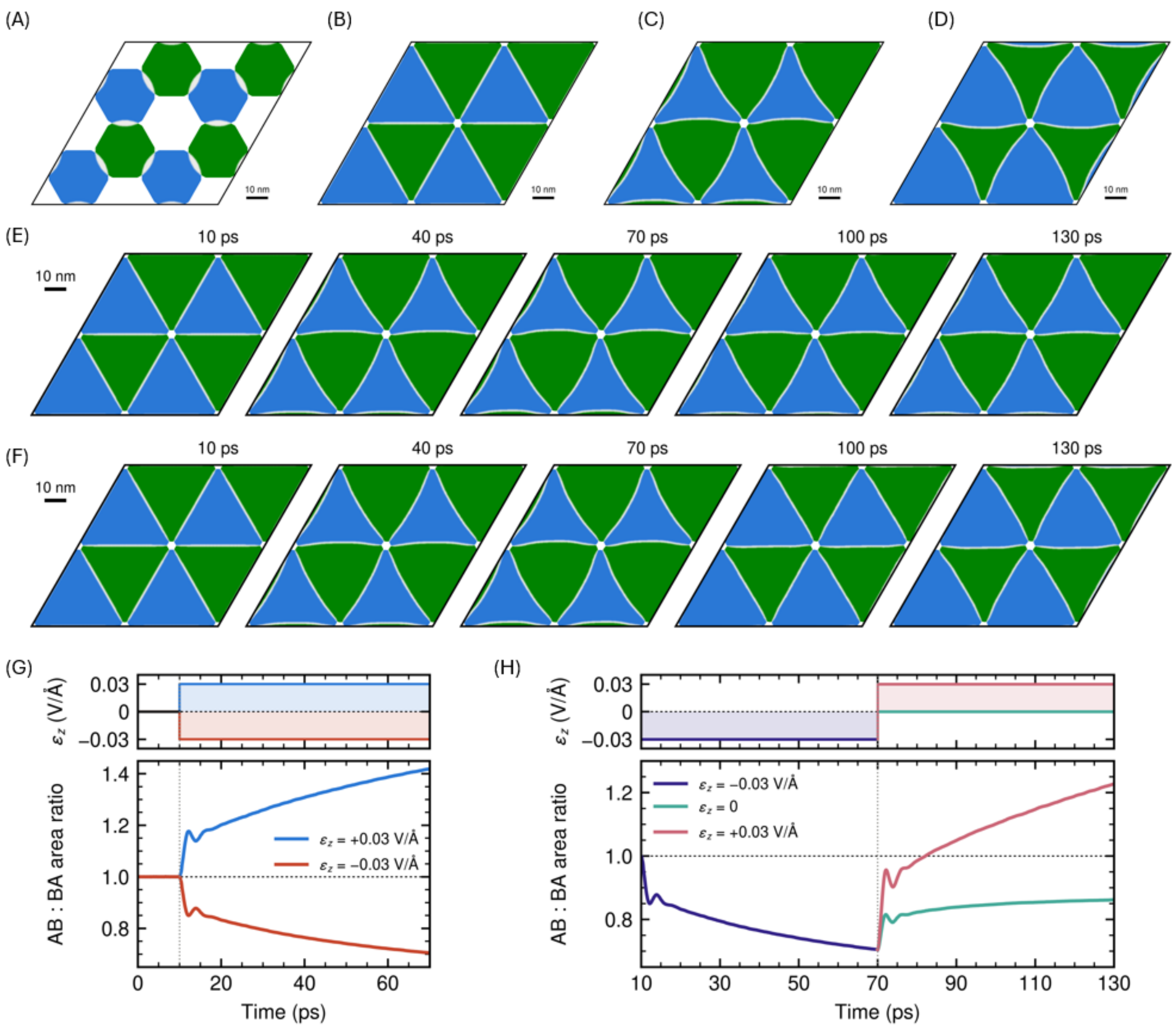

Figure 4. Reconstruction and electric-field response of triangular moiré domains in bilayer h-BN. (A) Stacking map of the unreconstructed supercell generated by a relative rotation of 0.336°. (B) Triangular domain network after 10 ps of zero-field relaxation. (C,D) Stacking maps after a further 60 ps under $\varepsilon_z = -0.03$ and +0.03 V Å$^{-1}$, respectively. (E,F) Stacking-map evolution when $\varepsilon_z = -0.03$ V Å$^{-1}$ is applied from t = 10 to 70 ps and then changed to zero or +0.03 V Å$^{-1}$, respectively, until t = 130 ps. The snapshots are taken at t = 10, 40, 70, 100, and 130 ps. (G) AB/BA area ratio under constant fields of −0.03 and +0.03 V Å$^{-1}$ applied at t = 10 ps, with the field histories shown above. (H) AB/BA area ratio during field removal and reversal, corresponding to E and F, with the field change at t = 70 ps. Blue and green regions denote AB and BA domains, respectively, and white regions identify DWs and AA nodes.

The snapshots show that AB contraction under negative $\varepsilon_z$ (Figure 4C) develops through inward-curving boundaries, while positive $\varepsilon_z$ expands the AB regions and reduces their BA neighbors (Figure 4D). The vertices remain connected through the shared node pattern. Boundary curvature and junction deformation consequently accompany the area change. This field-driven reshaping of connected polar domains is consistent with observations in experiments.[1,5,17,29] The speed of an isolated shear wall therefore provides a local propagation scale, while the evolving network geometry determines the collective area response.

The trajectory also exhibits nonuniform deformation near AA nodes and field-dependent interlayer breathing within the polar domains. Under negative $\varepsilon_z$, the AB separation exceeds the BA separation, and this ordering reverses with the field sign. This behavior extends the registry-dependent breathing observed in the monodomain calculations to a spatially heterogeneous network. Local deformation and boundary motion together accommodate the field without requiring all wall segments to translate at the same speed.

The deformed network retains a restoring tendency after the driving field changes (Figure 4E,F,H). Removing the negative field produces a rapid rebound of R, followed by slower partial recovery. The AB area remains smaller than the BA area over the displayed zero-field interval. Reversing the field to +0.03 V $Å^{-1}$ gives a larger rebound and sustained AB expansion, carrying R through unity. The recovery at zero field is consistent with relaxation of the curved boundaries and distorted junctions. Reversing the field adds a bias favoring AB and carries the network beyond equal domain occupation. Thus, the area reached at a given field depends on the preceding deformation and the time available for structural recovery.

The sharp rebound and subsequent oscillation following field removal or reversal reflect rapid structural relaxation on a changed electric enthalpy landscape. The abrupt change in the field-dependent polarization energy shifts the potential energy surface, driving the previously deformed structure toward a new equilibrium configuration. This rapid relaxation redistributes the AB and BA areas through changes in wall curvature and junction geometry before slower domain growth develops. A similar response is observed in the straight-wall trajectories in Figure 2E, while the connected walls and AA nodes introduce additional relaxation modes in the triangular domain structure. Such behavior highlights the importance of explicitly including field–polarization coupling in the potential energy model. Approaches combining a field-independent machine-learning potential with separately predicted BEC forces can capture the immediate force change but cannot reproduce the field-dependent energy landscape and the resulting structural relaxation.

Figure 5A follows the AB/BA area ratio during a field sweep between −0.03 and +0.03 V $Å^{-1}$. The embedded stacking maps show that both AB and BA domains persist as their relative areas change. The forward and return branches differ slightly, while R remains close to unity as the field passes through zero. The sampled cycle therefore shows little remanent domain imbalance. In addition, there is no clear evidence of macroscopic ferroelectric switching in this connected network. This behavior resembles the response of the moiré domain network observed experimentally in twisted $WSe_2$, where network connectivity suppresses conversion into a single polar state.[29] AB and BA retain their local polarizations throughout the deformation, and their opposing contributions approximately cancel near equal occupation.

The difference between the forward and reverse sweeps in Figure 5A reflects the history-dependent structural evolution of the triangular domains. Each field step abruptly changes the electric enthalpy landscape, driving rapid relaxation of DWs and junctions. As the structural configuration evolves under successive field steps, the system follows different relaxation pathways during the forward and reverse sweeps. This leads to different AB and BA area ratios even at the same applied field.

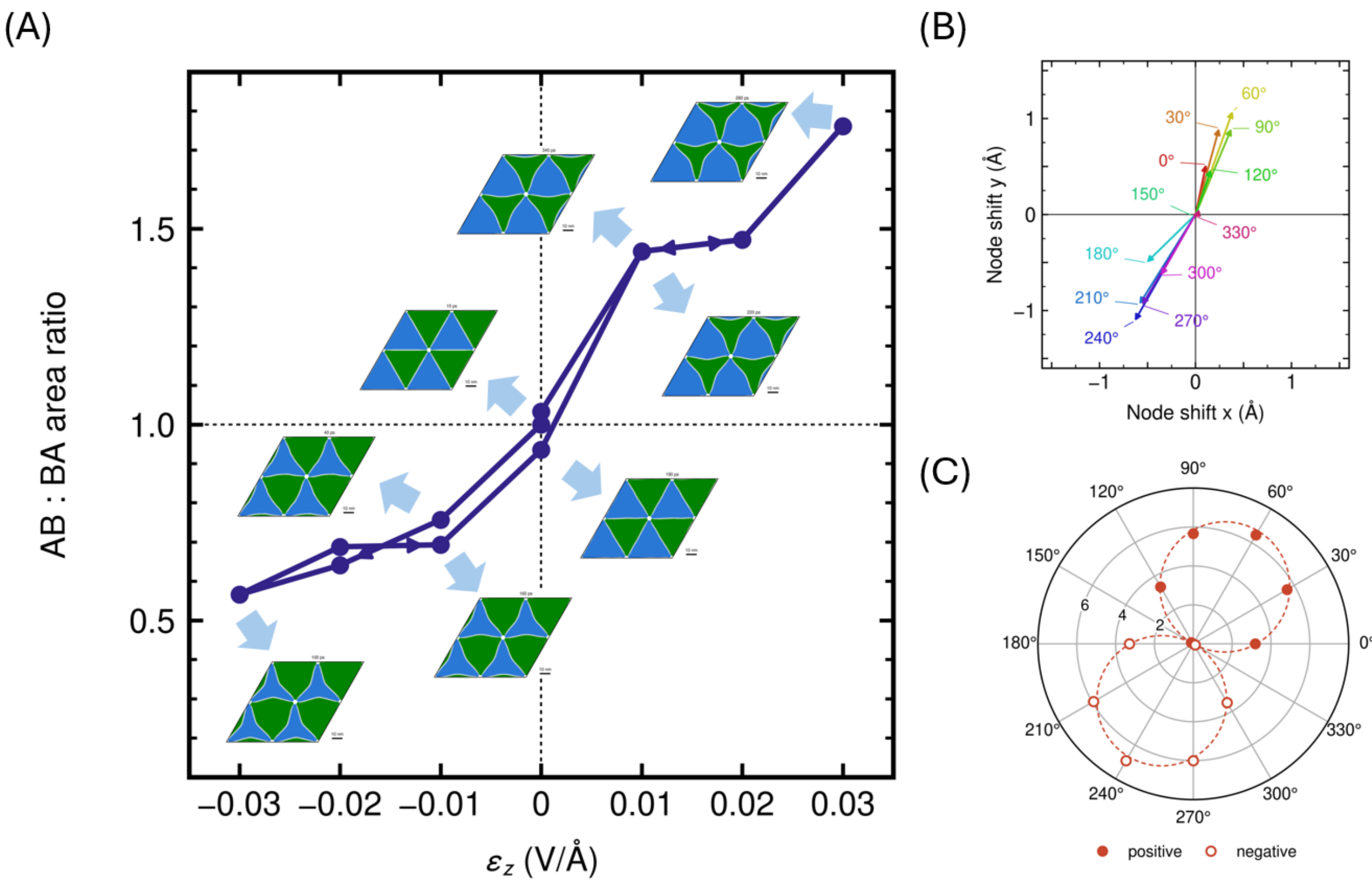

Figure 5. Domain-area response during a field sweep and directional sliding under an inclined electric field. (A) AB/BA area ratio versus $\varepsilon_z$ during a sweep between −0.03 and +0.03 V Å$^{-1}$. Arrows indicate the sweep direction, and eight stacking-map insets show representative configurations along the branches. (B) Displacements of the AA-node center for different in-plane-field angles at $\varepsilon_z$ = −0.03 V Å$^{-1}$ and $\varepsilon_\parallel$ = 0.01 V Å$^{-1}$. Arrow directions and lengths represent the directions and magnitudes of the node displacements in Å. (C) Signed projection of the relative in-plane layer translation onto the zigzag sliding axis at approximately −30° to the simulation-box +x axis, expressed in mÅ, under the same fields as in B. The relative translation is obtained from the change in the mean position of the upper layer relative to the lower layer. Both displacement readouts are evaluated at 60 ps relative to the corresponding trajectory with only $\varepsilon_z$ applied.

An auxiliary in-plane field produces node motion while leaving the AB/BA area ratio unchanged within the noise level across the sampled directions. With εz = −0.03 V Å$^{-1}$ fixed, an in-plane field of 0.01 V Å$^{-1}$ shifts the AA-node center along a nearly common axis (Figure 5B). Rotating the field changes the magnitude and sign of this displacement. Figure 5C shows the corresponding interlayer translation projected onto the zigzag sliding axis at approximately −30° relative to the simulation-box x axis. The displacement follows an approximately sinusoidal angular dependence, reaching 6.7 mÅ near θ = 60° and reversing sign under field reversal. Notably, the strongest response occurs when the applied field is nearly perpendicular to the atomic sliding direction. The large AA node displacement arises from moiré magnification, which scales approximately as the inverse twist angle in radians. At a twist angle of 0.336°, the amplification factor is about 170, so an interlayer translation of 6.7 mÅ produces a nearly perpendicular node displacement of approximately 1.1 Å, consistent with Figure 5B.

Motion near a junction can reshape the connected DWs without substantially changing the enclosed domain area. Consequently, the local wall displacement can exhibit a strong directional dependence on the in-plane electric field, while the overall AB/BA area ratio remains relatively insensitive to the field direction.

For an ideal threefold-symmetric network, the first-order contributions of three equivalent wall segments to the domain area rate cancel. Assuming equal geometrical weights c and velocity response vectors $\chi_j = \mathscr{R}(2\pi j/3)\chi_0$, where $\mathscr{R}$ denotes planar rotation, the field-induced area rate satisfies

$$\delta \dot{A}_{AB} = c \sum_{j=0}^{2} \chi_j \cdot \varepsilon_{ip} = 0 \tag{2}$$

In this ideal limit, the cancellation allows individual wall segments to respond directionally while their contributions to the total area rate compensate one another. We refer to this response as polarization-silent sliding. The layers translate and the moiré nodes move by approximately 1 Å, amplified geometrically by roughly 1/θ, while the polarization stays unchanged. Conversely, any perturbation that lifts the threefold equivalence, such as uniaxial strain, heterostrain, or nonuniform twist, should allow a finite net response. Connectivity is therefore a second gate, acting after local symmetry.

The triangular network thus converts electric-field coupling into a combination of domain-area change and local structural accommodation. The out-of-plane field selects which polar registry grows, while boundary curvature and shared nodes govern how that growth proceeds and recovers. An auxiliary in-plane field can induce relative layer sliding and a magnified translation of the moiré nodes with little change in the domain-area ratio. Across the monodomain, isolated-wall, and network calculations, the accessible structural motions determine how electric driving changes local registry and collective domain occupation. The three configurations studied here differ not in the underlying interatomic interactions but in the symmetry constraints on how the field couples to structure. Two gates operate sequentially. The first, local symmetry, determines whether a field-induced force exists along the sliding coordinate. The second, domain connectivity, determines whether local sliding sums to a change in net polarization. Closing the first gate requires large fields and spontaneous symmetry breaking; opening it permits wall motion at fields an order of magnitude weaker, and the second gate can decouple structural motion from polarization entirely. This perspective suggests two complementary design strategies. Breaking symmetry, through strain, ripples, or asymmetric networks, converts sliding into switching. Preserving network symmetry, by contrast, allows electrically driven structural actuation without perturbing the stored polarization. A comparison of the symmetry constraints and field-driven responses across the three configurations is provided in Section S9 and Table S2 of the SI.

## CONCLUSIONS

Our central finding is that electric-field control of sliding ferroelectrics is gated twice. Local symmetry determines whether the field can drive sliding, while domain connectivity determines whether sliding becomes switching.

In this study, we investigated electric-field-driven sliding and domain dynamics in bilayer h-BN across monodomain, isolated DW, and triangular moiré configurations using a machine-learning interatomic potential incorporating field-dependent electric enthalpy.

In an initially monodomain bilayer, where $C_{3v}$ symmetry forbids a net sliding force, switching near −0.45 V $Å^{-1}$ accompanies directional wrinkling, which lowers structural symmetry and opens a sliding route while the barrier along a constrained flat path remains finite. Pre-existing AB/BA DWs, whose reduced symmetry permits field-sliding coupling, respond at −0.03 V $Å^{-1}$, with sharply different propagation speeds set by their local shear and tensile character. At fixed out-of-plane bias, an additional in-plane field modulates wall velocity and width according to its orientation relative to the wall. The distinct angular responses of wall translation, broadening, and ionic displacement show that these processes cannot be treated as interchangeable measures of switching.

In triangular moiré networks, connected walls and AA junctions accommodate domain evolution through boundary curvature and collective deformation. After field removal or reversal, rapid structural readjustment precedes slower area evolution. An in-plane field also induces millångström-scale interlayer sliding and an approximately 1 Å displacement of moiré nodes without a resolvable change in AB/BA domain occupation. This polarization-silent sliding is a direct consequence of threefold connectivity, which cancels the contributions of individual walls to the net domain area. Lifting this symmetry, through uniaxial strain, heterostrain, or nonuniform twist, should therefore restore a finite polarization response to in-plane fields.

These results suggest two complementary design strategies. Breaking symmetry converts sliding into switching and lowers the field required for polarization reversal. Preserving network symmetry allows the lattice to be actuated electrically without disturbing the stored polarization. More broadly, distinguishing field-driven structural motion from net domain conversion provides a basis for their independent control.

## METHODS

### Model Preparation

Initial h-BN structures were obtained from the Materials Project database (mp-984).[33,34] Monolayer h-BN and AA′, AB, BA, and AA bilayers were constructed from the primitive cell. The bilayer training structures contained 64 atoms. Rigid interlayer translations of AA′ and AB bilayers along the $[1\bar{1}0]$ direction provided 300 configurations per sliding period. Moiré cells with $\sum 7$, $\sum 13$, $\sum 19$ commensurabilities were generated using InterfaceMaster.[35,36] A 15 Å out-of-plane vacuum gap was introduced using VASPKIT.[37]

First-principles reference calculations were performed using the Vienna Ab initio Simulation Package (VASP) with a plane-wave cutoff of 750 eV.[38,39] The nonlocal vdW-DF2-B86R exchange-correlation functional was used for structural relaxation and the energy and force calculations.[40] Finite-temperature configurations of AA′, AB, and BA bilayers were sampled using VASP on-the-fly machine-learning molecular dynamics at different temperatures.[41,42] Sampling was performed in the canonical (NVT) ensemble with a Langevin thermostat. BEC tensors were calculated using density functional perturbation theory (DFPT).[43,44] These calculations used the Perdew-Burke-Ernzerhof (PBE) functional and the projector augmented-wave (PAW) method.[15,45,46]

An h-BN-specific MACE-Field potential was obtained by fine-tuning against reference energies, atomic forces, and BEC tensors.[31] The dataset contained 6089 structures from the finite-temperature trajectories and interlayer sliding configurations. The data were stratified by stacking type and divided into training and test sets in an 8 to 2 ratio.

**Molecular Dynamics**

Production molecular dynamics (MD) simulations were performed in LAMMPS using the fine-tuned MACE-Field potential.[47] The force on atom i was obtained from the coordinate gradient of the field-dependent electric enthalpy at the instantaneous applied field

$$F_{i\alpha}(\mathbf{R}, \boldsymbol{\varepsilon}) = -\frac{\partial \mathcal{H}(\mathbf{R}, \boldsymbol{\varepsilon})}{\partial R_{i\alpha}} \tag{3}$$

Here, $\mathcal{H}$ denotes the electric enthalpy, $\mathbf{R}$ represents the atomic coordinates, and $\boldsymbol{\varepsilon}$ is the applied electric field. The coupling between the electric field and the BECs is detailed in Section S2 of the Supporting Information.

Monodomain simulations used a bilayer h-BN supercell containing 2304 atoms in the NVT ensemble with a Nosé-Hoover thermostat at 20, 50, 100, 200, and 300 K. Starting from zero, the out-of-plane field was cycled between extrema of −1.2 and +1.2 V Å$^{-1}$. Field magnitudes from 0.30 to 0.80 V Å$^{-1}$ were sampled in increments of 0.05 V Å$^{-1}$. Each field level was maintained for 12 ps, and the integration time step was 1.0 fs.

Rectangular DW cells and triangular moiré networks were simulated in the isothermal-isobaric (NPT) ensemble with a Langevin thermostat at a target temperature of 0 K. Cell relaxation was restricted to the in-plane x and y directions, with the out-of-plane cell dimension held fixed. At this target temperature, the thermostat introduces no thermal random force, allowing the field-driven structural response to be resolved without imposed thermal noise. The integration time step was 1.0 fs. Inclined fields were applied by adding the prescribed x and y components to the out-of-plane field.

## ASSOCIATED CONTENT

### Supporting Information

In preparation.

## DATA AVAILABILITY

The data and code supporting this study are available from the corresponding author upon request.

## AUTHOR INFORMATION

### Corresponding Author

Teruyasu Mizoguchi. Email teru@iis.u-tokyo.ac.jp.

### Author Contact Information

Yinan Wang. Email ynwang20020718@g.ecc.u-tokyo.ac.jp.

### Notes

The authors declare no competing financial interest.

## ACKNOWLEDGMENT

Y. Wang acknowledges support from the MERIT-WINGS program at the University of Tokyo. This work was supported in part by MEXT as "Next-Generation Computational Science Global

Reach & Innovation Program Japan" (Grant No. JPMXP1920260306). The authors thank Po-yen Chen (TSMC) for helpful discussions. Computations were performed using the GENKAI supercomputer at the Research Institute for Information Technology, Kyushu University, and the supercomputer Fugaku provided by the RIKEN Center for Computational Science through the HPCI System Research Projects.